\documentclass[a4paper, 12pt]{article}

\usepackage{a4wide}

\usepackage[T1]{fontenc}
\usepackage{lmodern}
\usepackage{textcomp}
\usepackage{microtype}

\usepackage[english]{babel}
\usepackage[autostyle, english = american]{csquotes}
\MakeOuterQuote{"}

\usepackage{amsmath}
\usepackage{mathtools}
\usepackage{upgreek}
\usepackage{units}
\usepackage{slashed}
\usepackage{physics}

\usepackage{subcaption}
\usepackage{booktabs}
\usepackage{graphicx}

\graphicspath{{./Figures/}}

\newcommand{\GeV}{\rm GeV}

\usepackage[sorting = none,style = numeric-comp,backend = bibtex, giveninits=true, bibencoding=utf8]{biblatex}
\bibliography{all}

\usepackage[hidelinks]{hyperref}

\makeatletter
\g@addto@macro\bfseries{\boldmath}
\makeatother

\usepackage{authblk}

\title{Probing the Seesaw Mechanism and Leptogenesis with the International Linear Collider\footnote{Talk presented at the International Workshop on Future Linear Colliders (LCWS2017), Strasbourg, France, 23-27 October 2017. C17-10-23.2.}
}

\author[1,2]{Stefan Antusch}
\author[1]{Eros Cazzato}
\author[3,4]{Marco Drewes}
\author[5]{Oliver Fischer}
\author[6]{Bj\"orn Garbrecht}
\author[6,2,4]{Dario Gueter}
\author[6,4]{Juraj Klaric}

\affil[1]{Department of Physics, University of Basel, Klingelbergstr. 82, CH-4056 Basel, Switzerland}
\affil[2]{Max-Planck-Institut f\"ur Physik (Werner-Heisenberg-Institut), \protect\\F\"ohringer Ring 6, 80805 M\"unchen, Germany}
\affil[3]{Centre for Cosmology, Particle Physics and Phenomenology, \protect\\Universit\'e catholique de Louvain, Louvain-la-Neuve B-1348, Belgium}
 \affil[4]{Excellence Cluster Universe, Boltzmannstr. 2, D-85748, Garching, Germany}
 \affil[5]{Institute for Nuclear Physics, Karlsruhe Institute of Technology, \protect\\Hermann-von-Helmholtz-Platz 1, D-76344 Eggenstein-Leopoldshafen, Germany}
\affil[6]{Physik Department T70, Technische Universit\"at M\"unchen,\protect\\James Franck Stra\ss e 1, D-85748 Garching, Germany}

\date{}

\begin{document}

\maketitle

\begin{abstract}
We investigate the potential of the International Linear Collider (ILC) to probe the mechanisms of neutrino mass generation and leptogenesis within the minimal seesaw model.
Our results can also be used as an 
estimate for the potential of a Compact Linear Collider (CLIC).
We find that heavy sterile neutrinos that simultaneously explain both, the observed light neutrino oscillations and the baryon asymmetry of the universe, can be found in displaced vertex searches at ILC. We further study the precision at which the flavour-dependent active-sterile mixing angles can be measured. The measurement of the ratios of these mixing angles, and potentially also of the heavy neutrino mass splitting, can test whether minimal type I seesaw models are the origin of the light neutrino masses, and it can be a first step towards probing leptogenesis as the mechanism of baryogenesis. Our results show that the ILC can be used as a discovery machine for New Physics in feebly coupled sectors that can address fundamental questions in particle physics and cosmology.
\end{abstract}

\paragraph{Neutrino masses and New Physics.} Neutrino flavour oscillations clearly indicate that neutrinos have masses. They are the only established piece of evidence for the existence of physics beyond the Standard Model (SM) of particle physics that has been observed in the laboratory.
Unveiling the origin of neutrino masses may therefore provide an important key to understand how the SM may be embedded into a more fundamental theory of nature.
In addition, it may also shed light on one of the deepest questions in cosmology, the \emph{baryon asymmetry of the universe} (BAU), i.e., the tiny excess $\sim 10^{-10}$ \cite{Ade:2015xua} of matter over antimatter in the early universe that formed the origin of the baryons we find today after mutual annihilation of all other particles and antiparticles, cf. e.g. \cite{Canetti:2012zc} for a discussion.
If CP-violation in the lepton sector is responsible for generating the BAU, then detailed studies of the \emph{neutrino portal} may pave the way for an understanding of the baryogenesis mechanism.
The International Linear Collider (ILC) is an excellent tool to study the neutrino portal during both, the Z-pole and high energy runs and can therefore be used as a discovery machine for New Physics in feebly coupled sectors that can address fundamental questions in particle physics and cosmology.
As an example we in the following study the type-I seesaw \cite{Minkowski:1977sc,GellMann:1980vs,Mohapatra:1979ia,Yanagida:1980xy,Schechter:1980gr,Schechter:1981cv}.

\paragraph{Low Scale Seesaw.} If the neutrino masses are at least partially generated by the Higgs mechanism in the same way as the masses of all other fermions in the SM, then this requires the existence of \emph{right handed neutrinos} $\nu_R$ to form a Dirac mass term $\bar{\nu_L}m_D\nu_R$. Here $\nu_L=(\nu_e,\nu_\mu,\nu_\tau)^T$ and $m_D=vY^\dagger$, $v$ is the Higgs field value and $Y$ is a $n_s \times 3$ matrix of Yukawa couplings.
$n_s$ is the number of right handed neutrino flavours, which must at least equal the number of non-zero light neutrino masses $m_i$ if the $\nu_R$ are the sole source of light neutrino masses.
The most general renormalisable Lagrangian that can be constructed from $\nu_R$ and SM fields reads
\begin{align}
\label{eq:Lagrangian}
\mathcal{L}=\mathcal{L}_{\rm SM}
+{\rm i} \overline{\nu_{{\rm R} i}}\partial\!\!\!/\nu_{{\rm R} i}
-\frac{1}{2}(\overline{\nu_{{\rm R} i}^c}M_{ij}\nu_{{\rm R} j} +  \overline{\nu_{{\rm R} i}}M_{ji}^*\nu_{{\rm R} j}^c)-Y_{ia}^*\overline{\ell_a}\varepsilon\phi \nu_{{\rm R} i}
-Y_{ia}\overline{\nu_{{\rm R} i}}\phi^\dagger \varepsilon^\dagger \ell_a.
\end{align}
Here $\ell_a$ with $a=e,\mu,\tau$ are the SM lepton doublets  and $\phi$ is the Higgs field. The superscript $c$ denotes charge conjugation, and $\varepsilon$ is the antisymmetric SU(2)-invariant tensor with the convention $\varepsilon_{12}=1$.
The $\nu_R$ can have a Majorana mass term $M$ with eigenvalues $M_i$ because they are gauge singlets.
The magnitude of the $M_i$ is unknown and may vary over many orders of magnitude, with different implications for particle physics, astrophysics and cosmology, see e.g. \cite{Drewes:2013gca}.
For $M_i\gg m_i$ the \emph{seesaw mechanism} is at work, and one can expand all quantities in the entries of the flavour matrix $\theta=m_D M^{-1}=v Y^\dagger M^{-1}$. The light neutrino mass matrix at second order in $\theta$ is $m_\nu=m_D M^{-1} m_D^T = \theta M \theta^T$.
The physical mass eigenstates can be described by the Majorana spinors
\begin{align}\label{LightMassEigenstates}
\nu_i=\left[ V_\nu^{\dagger}\nu_{\rm L}-U_\nu^{\dagger}\theta\nu_{\rm R}^c + V_\nu^{T}\nu_{\rm L}^c-U_\nu^{T}\theta\nu_{\rm R} \right]_i \ , \
N_i=\left[V_N^\dagger\nu_{\rm R}+\Theta^{T}\nu_{\rm L}^{c} +  V_N^T\nu_{\rm R}^c+\Theta^{\dagger}\nu_{\rm L}\right]_i\,.
\end{align}
Here
$V_\nu= (1-\frac{1}{2}\theta\theta^{\dagger})U_\nu$
is the matrix that diagonalises $m_\nu$, with $U_\nu$ its unitary part, while
$V_N =(1-\frac12 \theta^T\theta^*)U_N\simeq(1-\frac12 \theta^T\theta^*)$
diagonalises the heavy neutrino mass matrix
$M_N=M + \frac{1}{2}(\theta^{\dagger} \theta M + M^T \theta^T \theta^{*})$.
The $\nu_i$ can be identified with the familiar light neutrinos with masses $m_i$.
The existence of the additional heavy neutrinos $N_i$ is a prediction of the seesaw mechanism. The $N_i$ interact with a weak interaction that is suppressed by the mixing angles $\Theta_{ai}=(\theta U_N^*)_{ai}\simeq\theta_{ai}$ and via their Yukawa couplings to the physical Higgs particles.

\paragraph{Leptogenesis.}
The Yukawa interactions $Y_{ia}$ in general violate the charge-parity (CP) symmetry,
allowing the heavy neutrinos to generate a matter-antimatter asymmetry amongst leptons in the early universe \cite{Fukugita:1986hr}, which can be transferred into a baryon asymmetry via sphalerons processes \cite{Kuzmin:1985mm}. This process is called \emph{leptogenesis} and provides an elegant explanation for the observed BAU.\footnote{Various aspects of leptogenesis have recently been reviewed in refs.~\cite{Dev:2017trv,Drewes:2017zyw,Dev:2017wwc,Biondini:2017rpb,Chun:2017spz,Hagedorn:2017wjy}.}
The $N_i$ may therefore be the common origin on neutrino masses and baryonic matter in the universe.\footnote{$N_i$ with sufficiently small $|\theta_{ai}|$ are also a viable DM candidate \cite{Dodelson:1993je,Shi:1998km}, see \cite{Adhikari:2016bei} for a recent review.
However, those $N_i$ that compose the DM cannot make a significant contribution to the generation of light neutrino masses and leptogenesis because their Yukawa couplings are constrained to be tiny in order to make them sufficiently long lived. This does of course not exclude the possibility that different $N_i$ flavours in the same model can play the two different roles: One of them may compose the Dark Matter while the two (or more) others  can explain the neutrino masses and the BAU. This possibility has been proposed in the $\nu$MSM \cite{Asaka:2005an,Asaka:2005pn}, its feasibility in that model was show in refs.~\cite{Canetti:2012vf,Canetti:2012kh}.
}
In the mass range $M_i <$ TeV that is accessible to collider experiments, leptogenesis can proceed in two different ways. For $M_i$ above the electroweak scale, the BAU can be generated during the freeze-out and decay of the $N_i$ \cite{Pilaftsis:2003gt} in the early universe ("freeze-out scenario"). For masses below the electroweak scale, the BAU can be generated in CP-violating oscillations \cite{Akhmedov:1998qx,Asaka:2005pn} and Higgs decays \cite{Hambye:2016sby} during their production ("freeze-in scenario").
We focus on the second possibility, which allows for an efficient production of $N_i$ at the ILC in  weak gauge boson decays.
The minimal implementation of this scenario with $n_s=2$ \cite{Shaposhnikov:2008pf,Canetti:2010aw,Asaka:2011wq,Canetti:2012vf,Canetti:2012kh,Shuve:2014zua,Hernandez:2015wna,Drewes:2016lqo,Drewes:2016jae,Hernandez:2016kel,Asaka:2016zib,Drewes:2016gmt,Asaka:2017rdj,Antusch:2017pkq}, its realisation within inverse and linear seesaw models \cite{Abada:2015rta,Abada:2017ieq} and
 the slightly more general case with $n_s=3$ heavy Majorana neutrinos \cite{Drewes:2012ma,Khoze:2013oga,Canetti:2014dka,Shuve:2014zua,Hernandez:2015wna,Drewes:2016lqo} have been studied by many authors.
Here we consider the minimal model with $n_s=2$, i.e., the smallest number of  $N_i$ that is required for consistency with light neutrino oscillation data.
This effectively also described leptogenesis in the $\nu$MSM.
In this scenario the $N_i$ must have quasi-degenerate masses
with $\upmu=|M_2-M_1|/(M_2+M_1)<0.1$ \cite{Antusch:2017pkq} to generate the observed BAU. For $n_s>2$ no degeneracy is required for leptogenesis \cite{Drewes:2012ma,Shuve:2014zua,Canetti:2014dka,Hernandez:2015wna}.

\begin{figure}
\begin{center}
\includegraphics[width=0.7\textwidth]{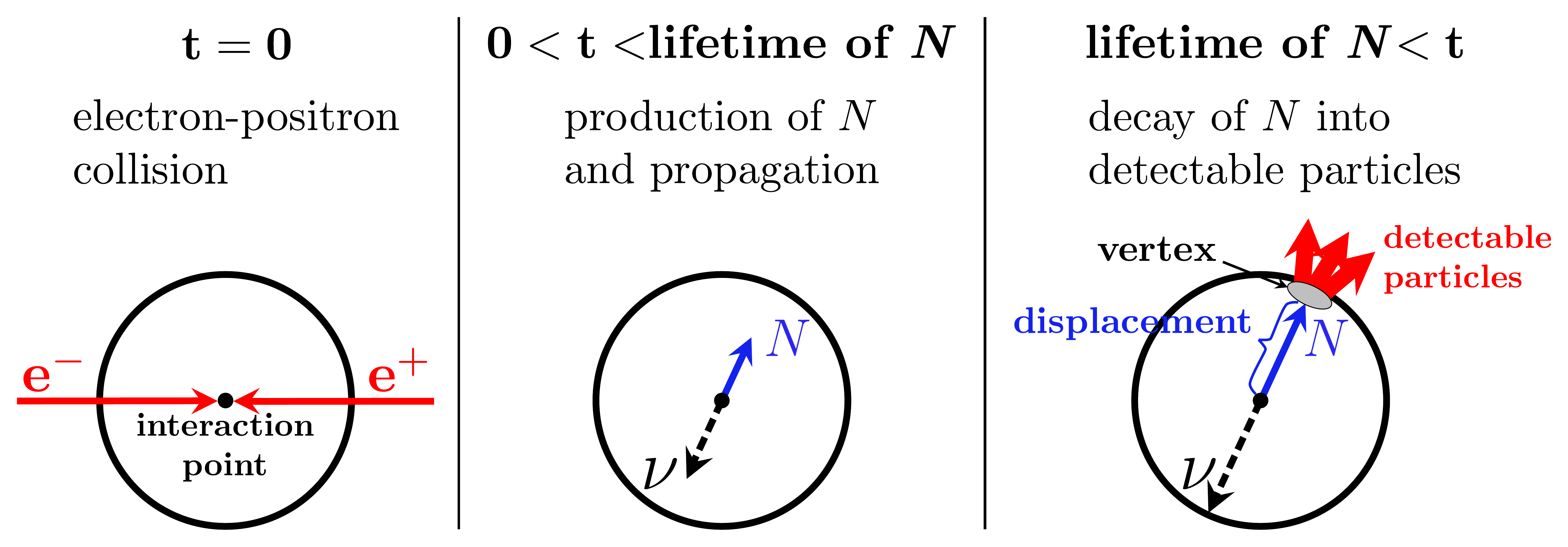}
\end{center}
\caption{Heavy neutrinos with masses below the electroweak scale are long lived particles with lifetimes $\propto U_i^{-2}M_i^{-5}$ \cite{Gorbunov:2007ak}. This leads to a displacement between the collision point and the decay vertex.}
\label{fig:displacedvertex}
\end{figure}

\paragraph{Searches at the ILC.}
High energy colliders provide the best tool to search for $N_i$ with masses above 5 GeV. For smaller masses, fixed target experiments like NA62  \cite{Lazzeroni:2017fza,CortinaGil:2017mqf,Drewes:2018gkc} or SHiP \cite{Anelli:2015pba,Alekhin:2015byh} are more sensitive.
An overview of possible signatures at different collider types can e.g. be found in ref.~\cite{Banerjee:2015gca,Deppisch:2015qwa,Antusch:2016ejd,Cai:2017mow}. At lepton colliders \cite{Antusch:2016vyf} displaced vertex searches turn out to provide the highest sensitivity for $M_i$ below the electroweak scale, cf. figure \ref{fig:displacedvertex}.\footnote{Previous studies suggest that the LHC cannot probe the parameter region where leptogenesis is possible in the minimal model with $n_s=2$ because the $U_{ai}^2$ required for leptogenesis \cite{Antusch:2017pkq} are too small to yield observable event rates~ \cite{Helo:2013esa,Izaguirre:2015pga,Gago:2015vma,Antusch:2017hhu}.
However, the possibility of a flavour asymmetric washout can make leptogenesis feasible for larger $U_{ai}^2$ that can be probed at the LHC for $n_s>2$ \cite{Canetti:2014dka}. Moreover, upgrades like MATHUSLA \cite{Chou:2016lxi} could increase the sensitivity of the LHC.
}

The event rates are determined by the quantities $U_{ai}^2=|\Theta_{ai}|^2$, which characterise the interaction strength of heavy neutrino $N_i$ with SM leptons $\ell_a$.
Since both, $m_i$ and $U_{ai}^2$ are $\propto \theta_{ai}^2$, it is in general difficult to make observable event rates consistent with the small neutrino masses \cite{Kersten:2007vk}.
It is, however, possible in a natural way if one assumes that the exact $B-L$ symmetry of the SM is approximately preserved by whatever New Physics the Lagrangian (\ref{eq:Lagrangian}) is embedded into \cite{Shaposhnikov:2006nn,Kersten:2007vk,Moffat:2017feq}.
This can e.g. be realised in "inverse seesaw" type scenarios \cite{Wyler:1982dd,Mohapatra:1986aw,Mohapatra:1986bd,Bernabeu:1987gr}, with a "linear seesaw" \cite{Akhmedov:1995ip,Akhmedov:1995vm}, in scale invariant models \cite{Khoze:2013oga} or the \emph{Neutrino Minimal Standard Model} $\nu$MSM \cite{Asaka:2005an,Asaka:2005pn,Shaposhnikov:2006nn}.
In the symmetric limit one observes $U_{a1}^2=U_{a2}^2$ and $M_1=M_2\equiv \bar{M}$. It is therefore instructive to introduce the quantities $U_a^2=\sum_i U_{ai}^2$, with $U_{ai}^2\simeq U_a^2/2$.
The symmetry automatically provides a natural explanation for the $\upmu\ll1$ required for leptogenesis.

If the collider is operated at at the Z-pole ($\sqrt{s}=90$ GeV), then $N_i$ with $M_i< 90$ GeV are primarily produced via $s$-channel exchange of on-shell Z bosons along with a SM neutrino $\nu_a$. At higher collision energies the
production through charged current interactions
dominates.
This has direct implications for the dependence of the production rates on the heavy neutrino flavour mixing pattern, i.e., the relative size of the $U_{ai}^2$ for fixed $U_i^2 = \sum_a U_{ai}^2$. The reason is that the production via $s$ channel $Z$ bosons is independent of the flavour of the associated neutrino $\nu_a$ (and therefore only depends on $U_i^2$), while the production via channel W boson exchange necessarily involves the electron flavour and is always proportional to $U_{e i}^2$.
The $N_i$ are detected via their weak decays into charged particles.
If all masses in the final state can be neglected, then the decay rate practically only depends on $U_i^2$.
Hence, in Z pole runs the total event rate is roughly determined by $M_i$ and $U_i^2$ alone, while it depends on the flavour mixing pattern for higher energy runs.
\begin{figure}
\begin{subfigure}{0.5\textwidth}
\includegraphics[width = \textwidth]{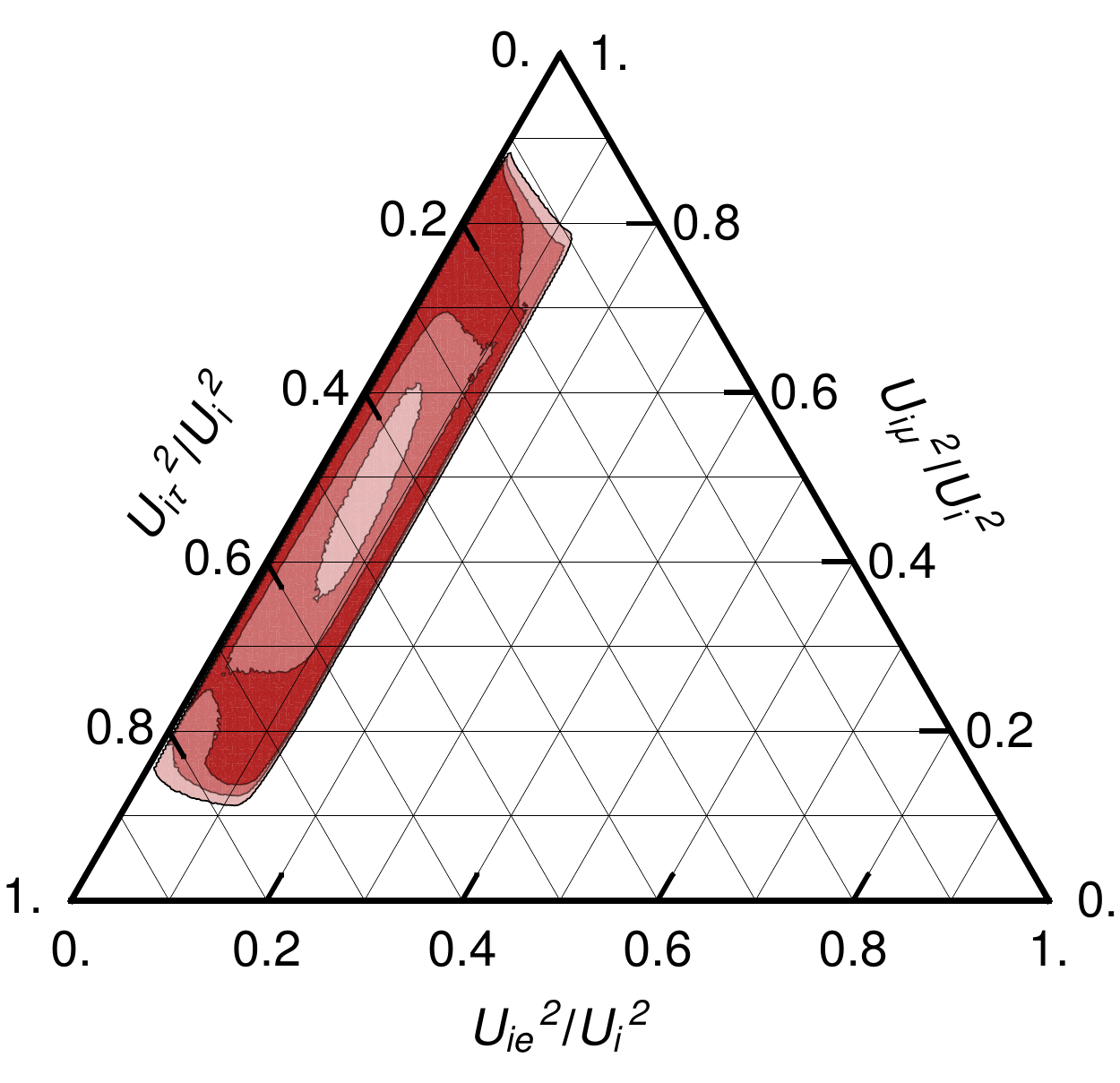}
\caption{normal ordering}
\label{fig:NOalpha}
\end{subfigure}
\begin{subfigure}{0.5\textwidth}
\includegraphics[width = \textwidth]{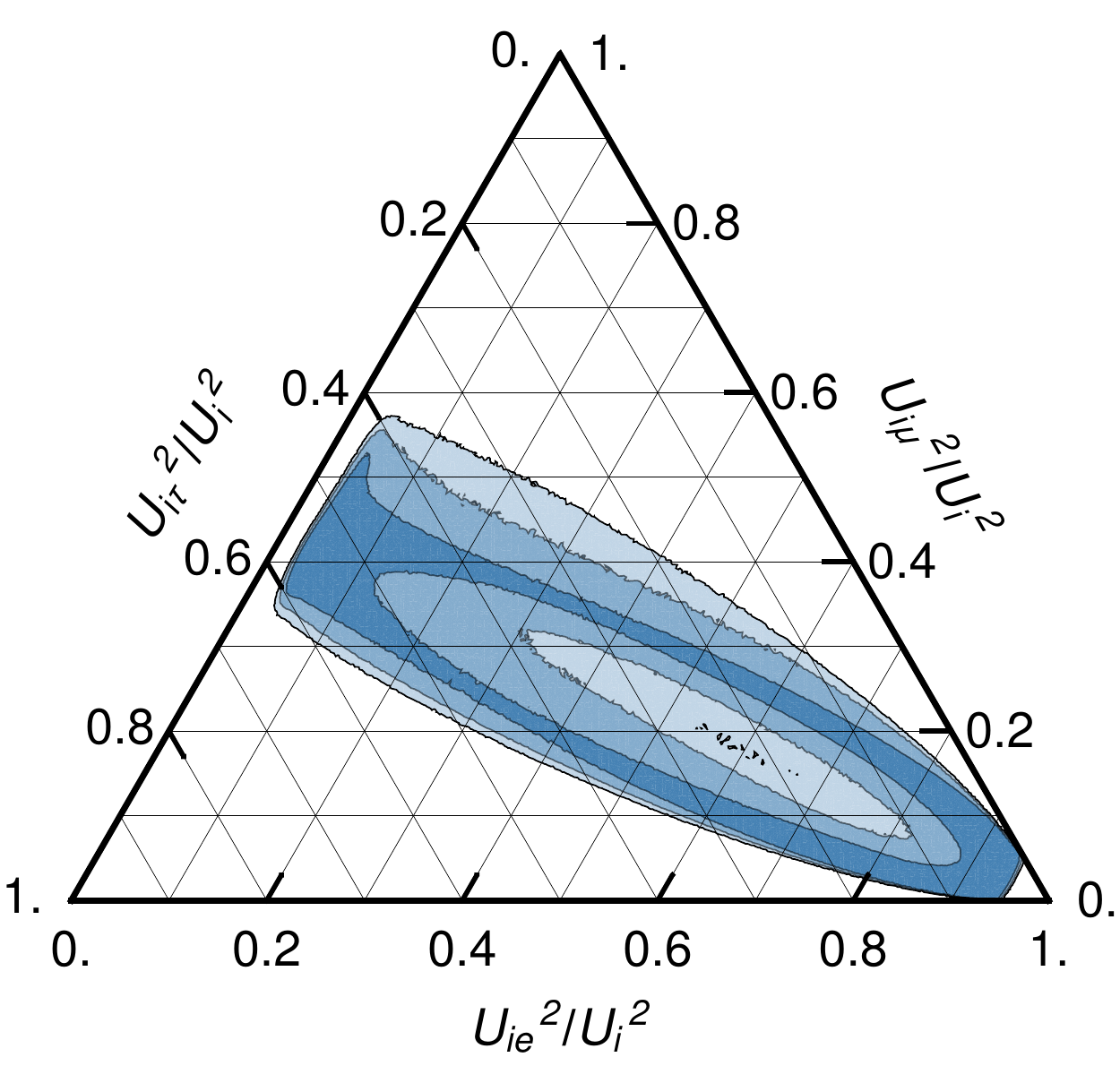}
\caption{inverted ordering}
\label{fig:NOsin}
\end{subfigure}
\caption{The different shades indicate the 1$\sigma$ (darkest), 2$\sigma$ and 3$\sigma$ (lightest) probability contours for the ratios $U_a^2/U^2$ for $n_s = 2$ that can be obtained from the NuFIT~3.1 global fit to neutrino oscillation data ~\cite{Esteban:2016qun, nufit}, assuming a flat prior for the unconstrained Majorana phase.
The results depend only mildly on the choice of this prior.
Figure taken from ref.~\cite{Drewes:2018gkc}.
\label{fig:Chi2NO}}
\end{figure}
The ratios $U_{ai}^2/U_i^2$ are strongly constrained by neutrino oscillation data \cite{Shaposhnikov:2008pf,Gavela:2009cd,Asaka:2011pb,Ruchayskiy:2011aa,Hernandez:2016kel,Drewes:2016jae,Drewes:2018gkc}.
In ref.~\cite{Drewes:2018gkc} it was shown that the combined data from neutrino oscillation experiments is sufficient to identify statistically preferred regions for the $U_{ai}^2/U_i^2$, cf. figure \ref{fig:Chi2NO}.
Hence, it is possible to define a "most optimistic" and a "most pessimistic" scenario for the high energy run within this allowed range.

\begin{figure}
        \centering
        \begin{tabular}{cc}
			\textbf{Normal Ordering} & \textbf{Inverted Ordering} \\\\
            \includegraphics[width=0.43\textwidth]{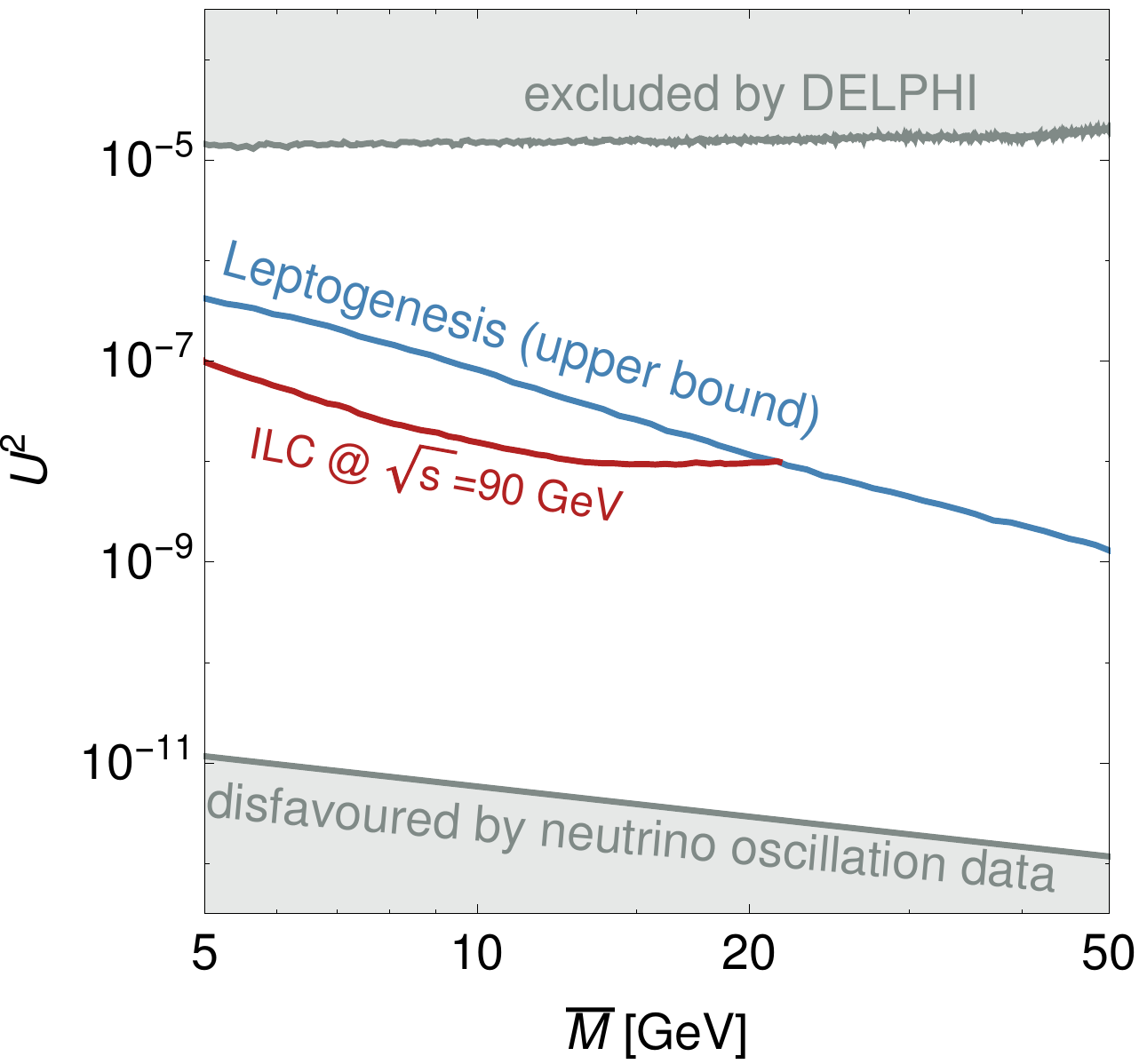} &
			\includegraphics[width=0.43\textwidth]{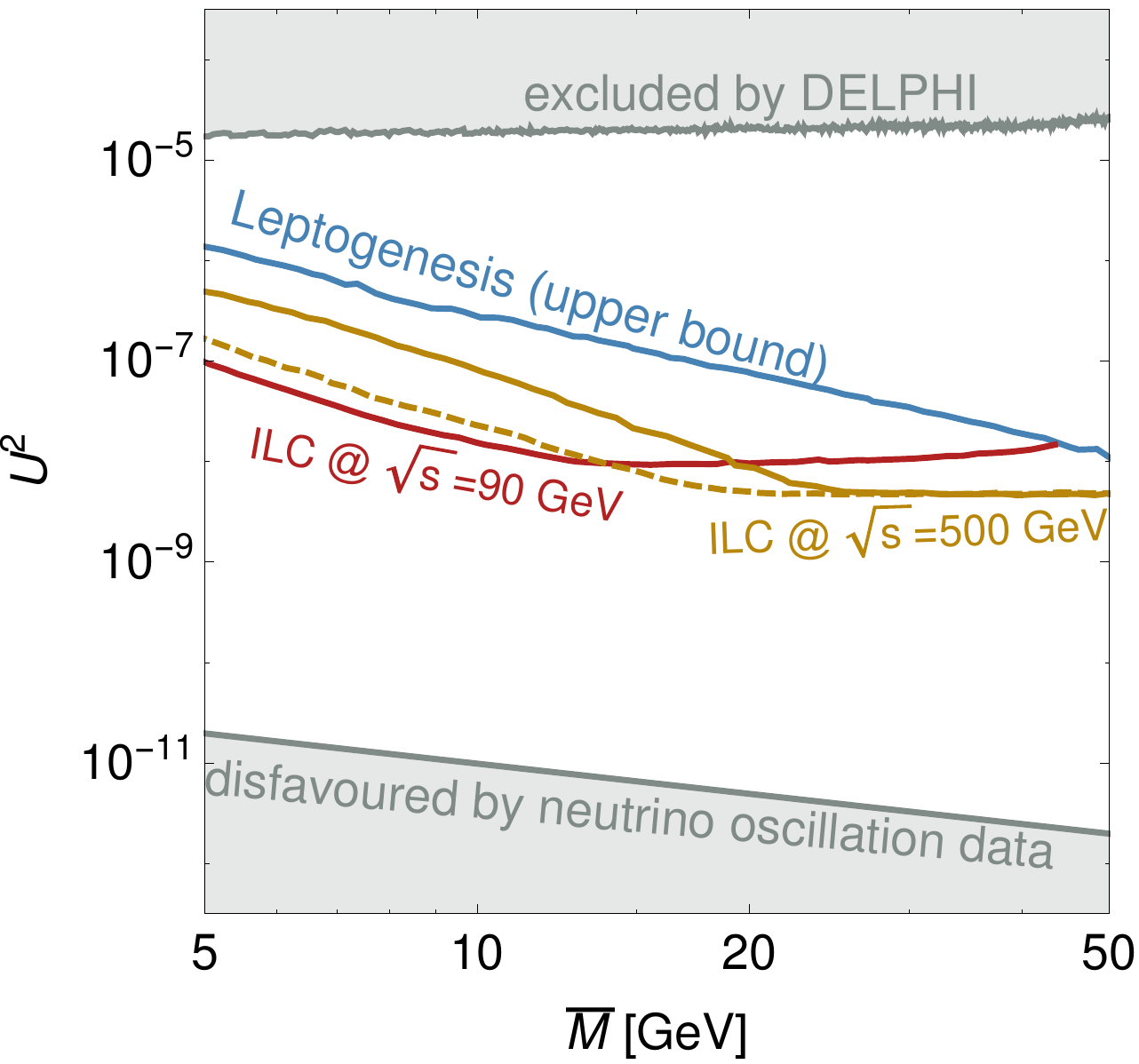}
        \end{tabular}
\caption{\label{fig:ILCcomparison}
The blue ``BAU'' line shows the largest possible mixings $U^2=\sum_i U_i^2$ for which the BAU can be generated in the seesaw model with $n_s=2$ for given $\bar{M}=|M_2+M_1|/2$.
The grey area is ruled out by the DELPHI experiment \cite{Abreu:1991pr,Abreu:1996pa} (on the top) and by neutrino oscillation data (at the bottom).
We display no lower bound on $U^2$ from leptogenesis because this constraint is weaker than that from neutrino oscillation data in this mass range.
The coloured lines mark the parameter regions in which the ILC experiment is expected to observe at least four displaced vertex events from $N_i$ with properties that are consistent with successful leptogenesis.
The orange lines show the regions accessible with $\sqrt{s}=500$ GeV and an integrated luminosity of $\mathcal{L}=0.1\,\text{ab}^{-1}$ and $\mathcal{L}=5\,\text{ab}^{-1}$, which depend on the relative size of the heavy neutrinos mixings $U_{a i}^2$ to individual SM flavours because the $N_i$ production is dominated by charged current interactions, which necessarily involve the mixing $U_{e i}^2$ with the electron flavour.
The solid and dashed lines correspond to the most optimistic and most pessimistic scenario consistent with light neutrino oscillation data. The lack of orange lines in the left panel is due to the suppression of $U_{e i}^2$ for normal ordering of light neutrino masses for $n_s=2$, cf. figure \ref{fig:Chi2NO}. This suppression is less efficient for $n_s>2$ \cite{Gorbunov:2013dta,Drewes:2015iva}.
The purple lines indicate the regions accessible with $\sqrt{s}$ at the Z pole, which only depend on the total $U_i^2$. Figure taken from ref.~\cite{Antusch:2017pkq}.
}
\end{figure}

\begin{figure}
        \centering
        \begin{tabular}{cc}
			\textbf{NO, ILC at $\sqrt{s}= 90 \,{\rm GeV}$} & \textbf{IO, ILC at $\sqrt{s}= 90 \,{\rm GeV}$} \\
            \includegraphics[width=0.5\textwidth]{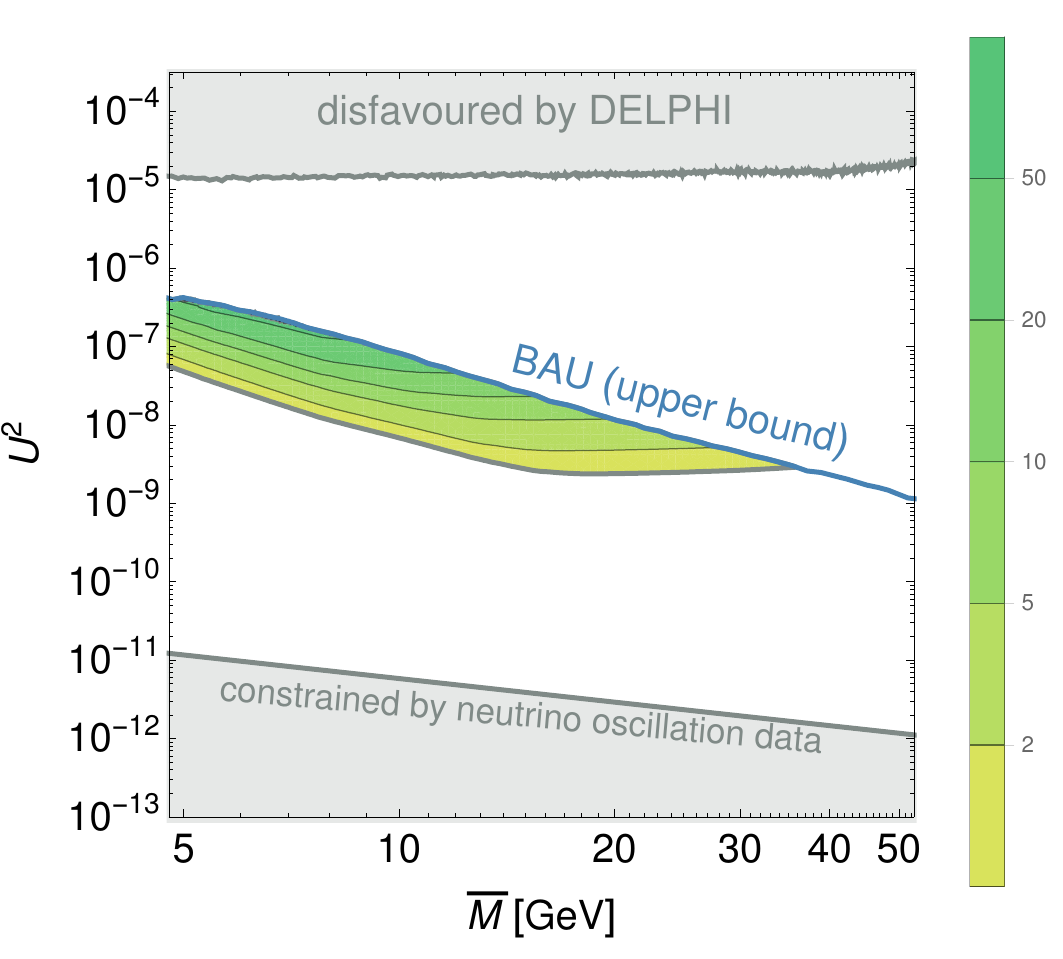} &
			\includegraphics[width=0.5\textwidth]{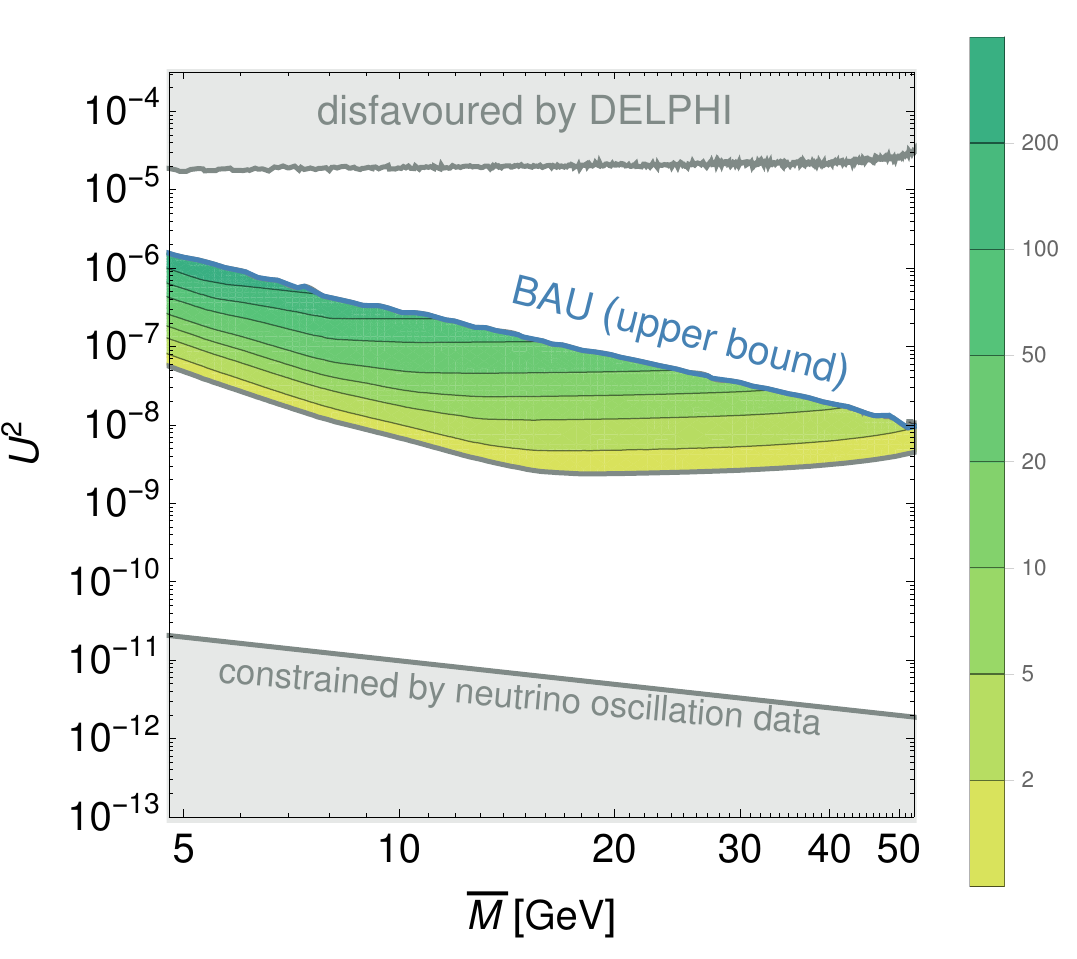}\\
			\textbf{NO, ILC at $\sqrt{s}= 500 \,{\rm GeV}$} & \textbf{IO, ILC at $\sqrt{s}= 500 \,{\rm GeV}$} \\

            \includegraphics[width=0.5\textwidth]{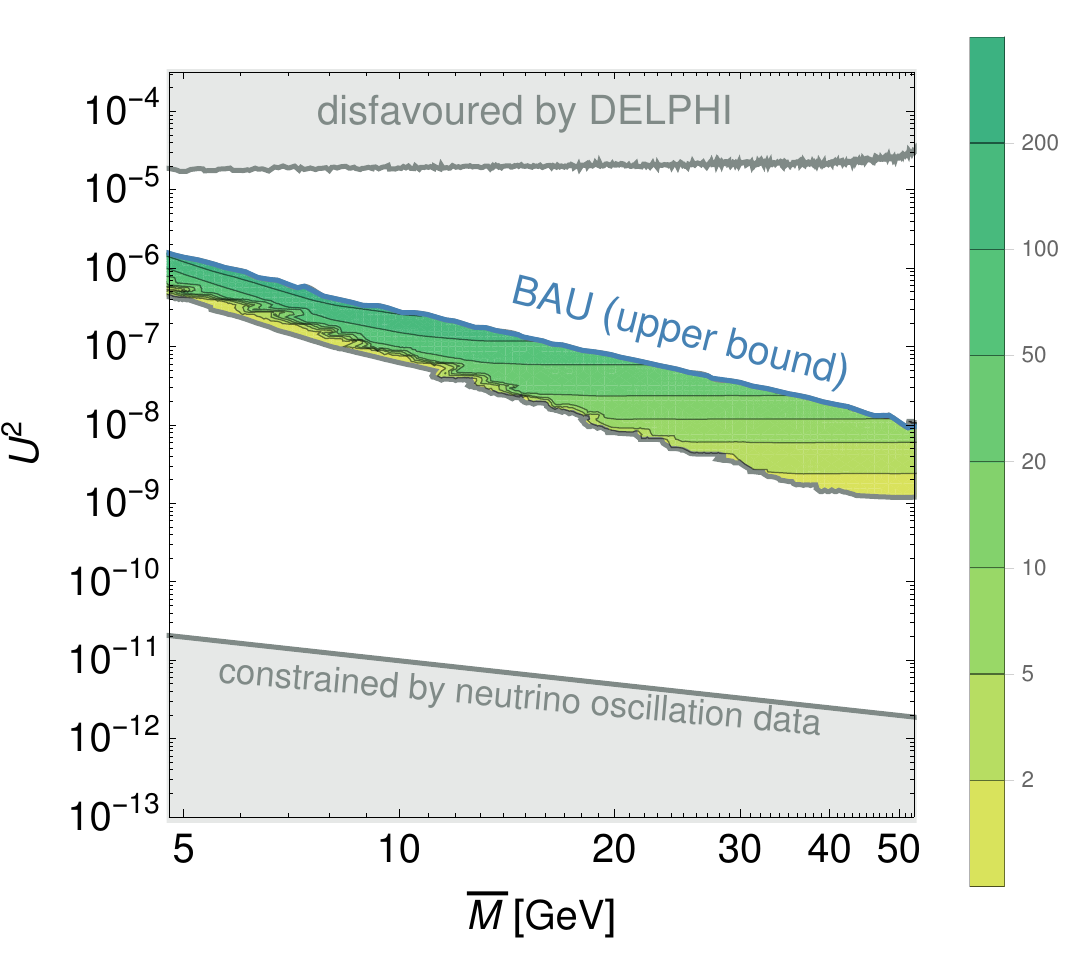}

             &
			\includegraphics[width=0.5\textwidth]{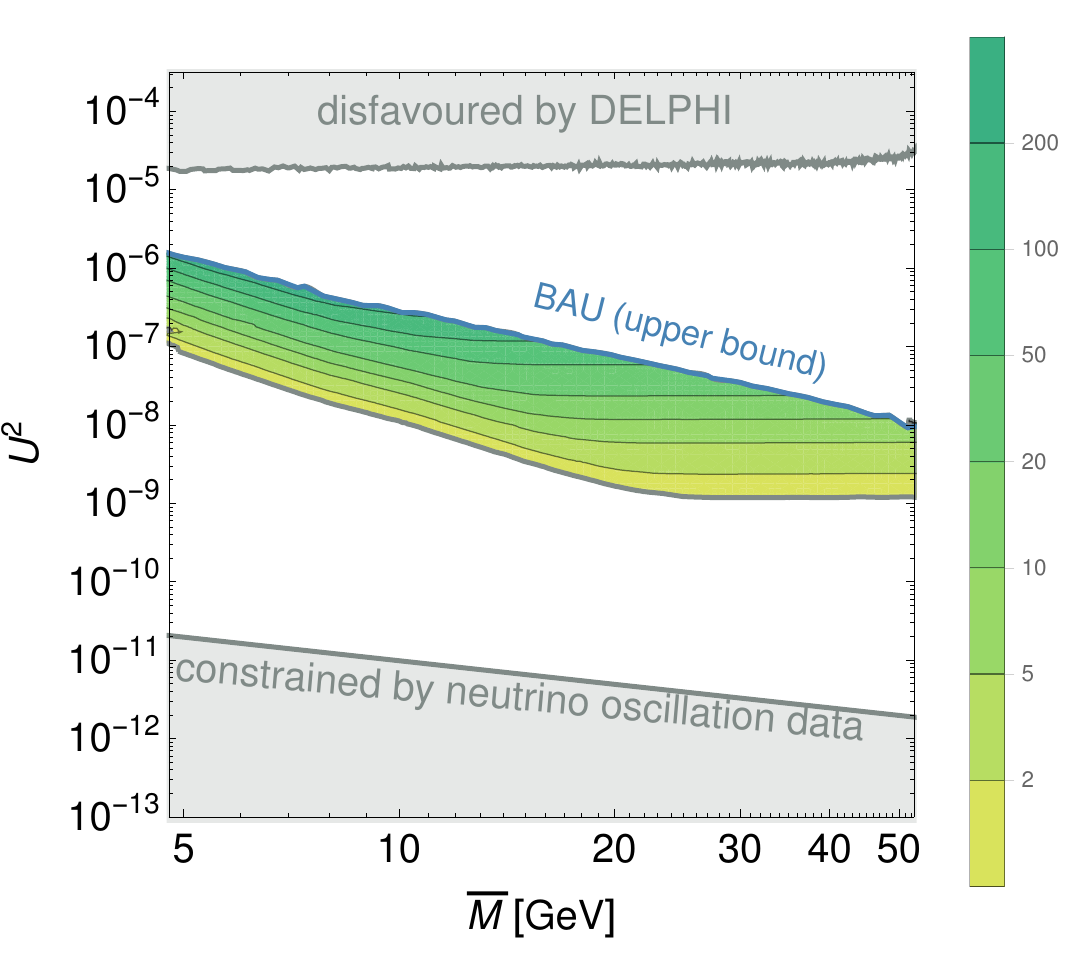}\\
			minimal number of events & maximal number of events\\
        \end{tabular}
\caption{\label{NumberOfEvents500}
The number of events expected to be seen 
at the ILC with $\sqrt{s}=90\,\GeV$ (upper panels) and $\sqrt{s}=500\,\GeV$ (lower panels) as a function of $M_i$ and $U_i^2$ for the most pessimistic (small $U_{ei}^2/U_i^2$, left panel) and most optimistic
 (large $U_{ei}^2/U_i^2$, right panel) consistent with neutrino oscillation data and successul leptogenesis (cf. figure~\ref{fig:Chi2NO}) with $n_s=2$ and IO. Note that the number can be much larger above the "BAU" line, where leptogenesis is not feasible for $n_s=2$, but the low scale seesaw mechanism can still provide an explanation for the observed neutrino oscillation. Moreover, leptogenesis is believed to be feasible for larger $U_i^2$ with $n_s>2$ \cite{Canetti:2014dka}.
Figure taken from ref.~\cite{Antusch:2017pkq}.
}
\end{figure}

The ILC sensitivity for the different cases is shown in figure~\ref{fig:ILCcomparison}. Figs~\ref{fig:TotalU2M_ILC_CECP} and \ref{NumberOfEvents500} show the expected number of events.
The numerical calculation of the cross section for the different discussed performance parameters of the considered colliders is done in \texttt{WHIZARD} \cite{Kilian:2007gr, Moretti:2001zz} by including initial state radiation and by including also a (L,R) initial state polarisation of (80\%,20\%) and beamstrahlung effects.

\begin{figure}
	\centering
	\begin{tabular}{cc}
		\textbf{Normal ordering} & \textbf{Inverted ordering}\\
\includegraphics[width=0.45\textwidth]{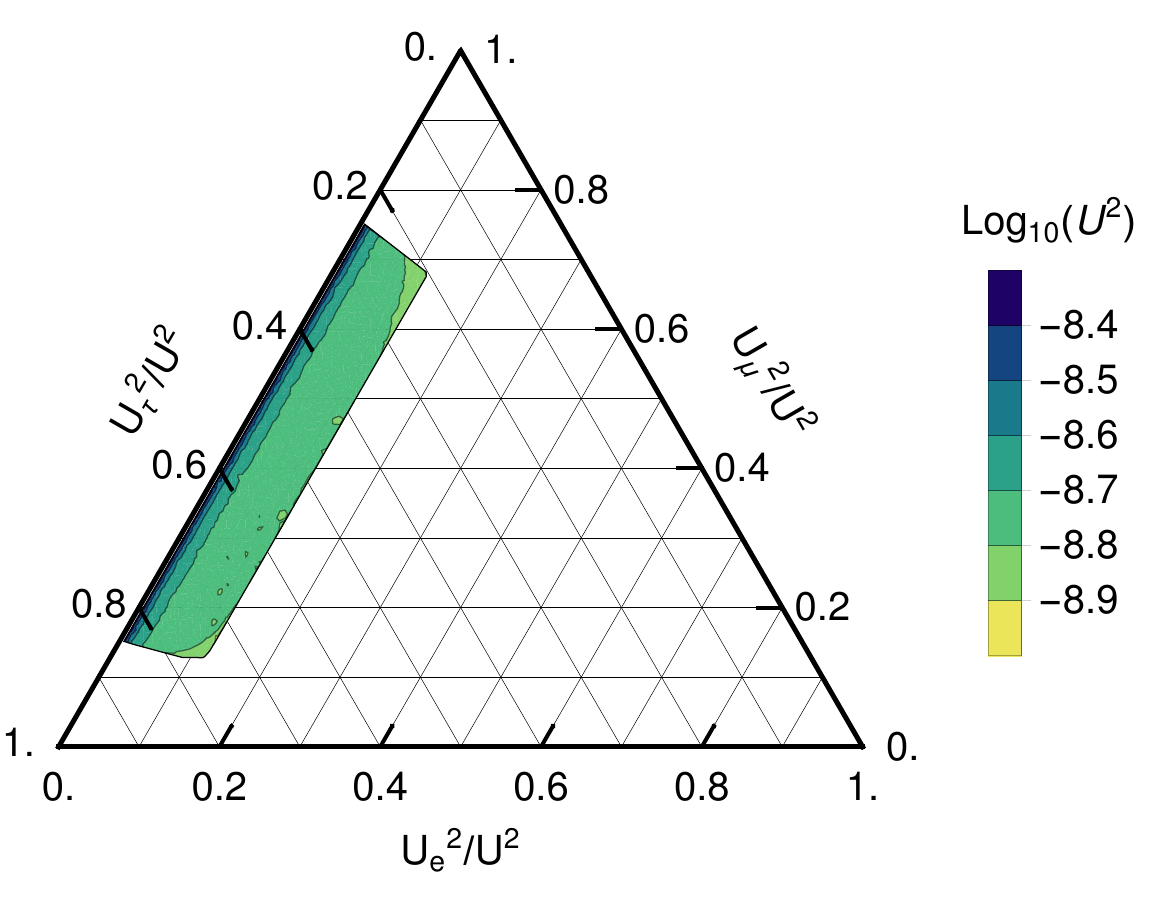}&
\includegraphics[width=0.45\textwidth]{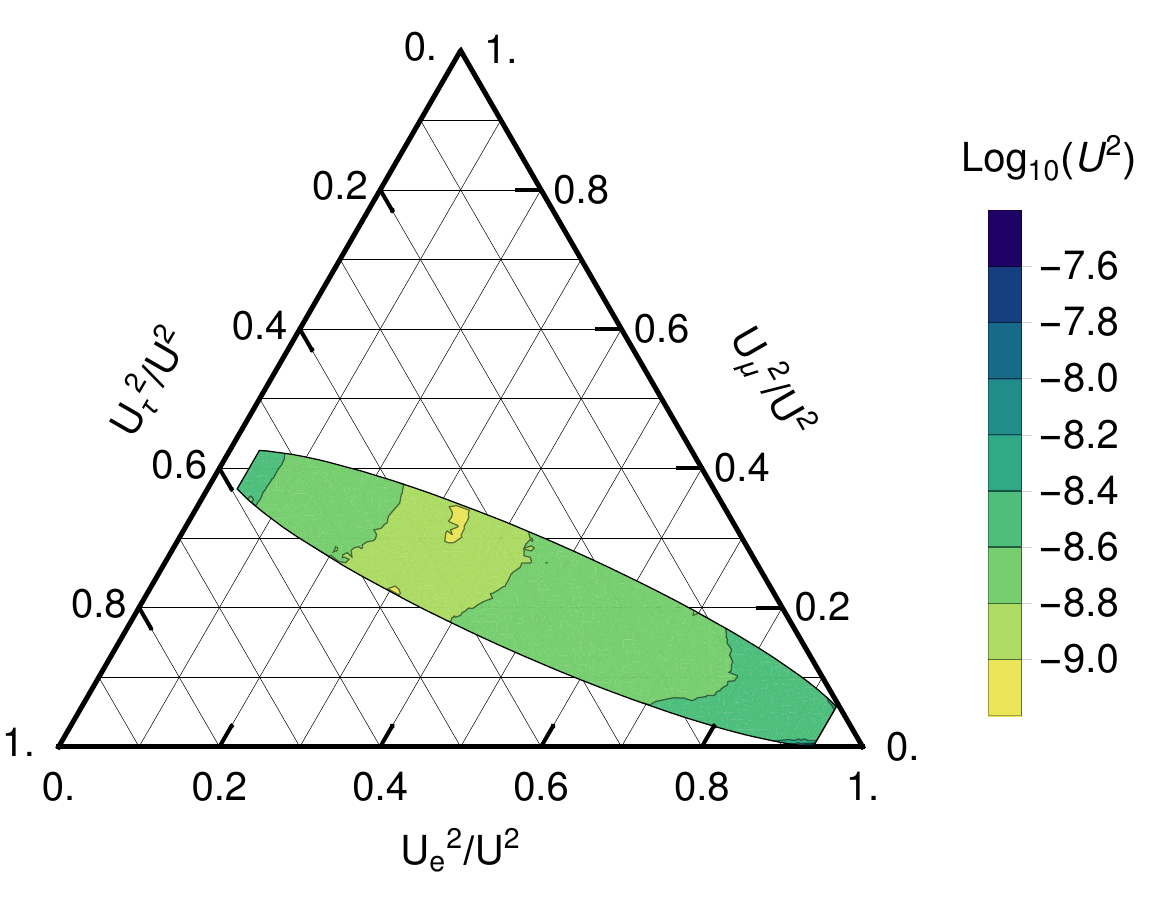}
\end{tabular}
	\caption{The region within the black lines is allowed by light neutrino oscillation data for $n_s=2$, cf. figure \ref{fig:Chi2NO}.
	The colour indicates the largest mixing angle $U^2$ that allows to produce the observed BAU for the cases of normal ordering (left) and inverted ordering (right) for right-handed neutrino with an average mass $\bar{M}=30\, \GeV$. The largest viable mixing angles are found in the case of a highly hierarchical flavour mixing pattern ($U^2_a\ll U^2$ for one of the flavours). This hierarchy allows to protect part of the asymmetries from the washout in the early universe even if $U^2$ is large enough that the heavy neutrinos come into equilibrium before sphalerons freeze out.
	The hierarchy between the smallest $U_{a i}^2$ and $U_i^2$ can be much larger for $n_s>2$, which makes leptogenesis feasible for larger $U_i^2$ \cite{Canetti:2014dka} and thereby improves the perspectives for an experimental discovery in comparison to the minimal $n_s=2$ scenario discussed here.
	 }
	\label{fig:triangleplt}
\end{figure}

\begin{figure}[t]
        \centering
        \begin{tabular}{ccc}
	\includegraphics[width=0.4\textwidth]{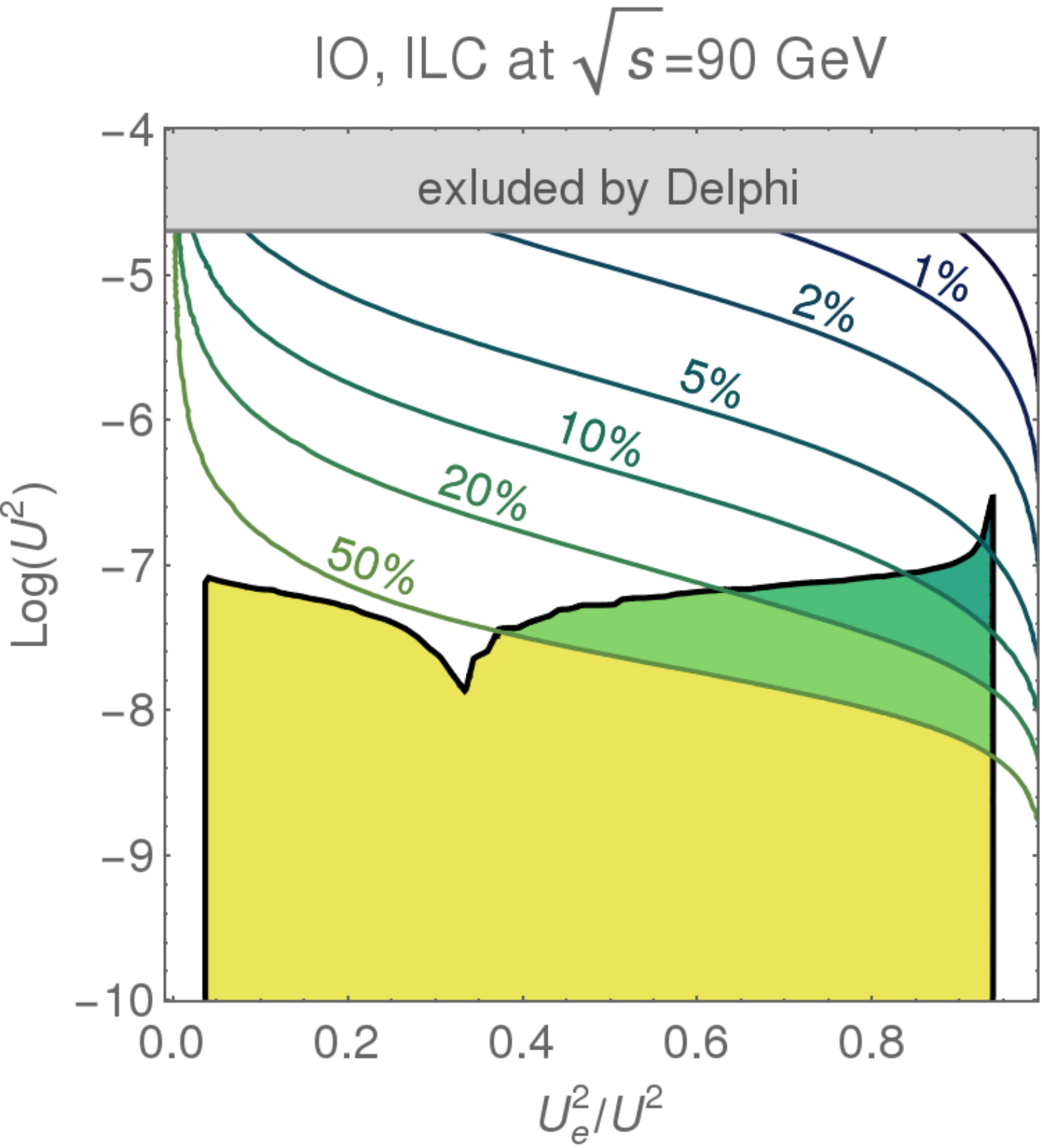}
			&
			\includegraphics[width=0.4\textwidth]{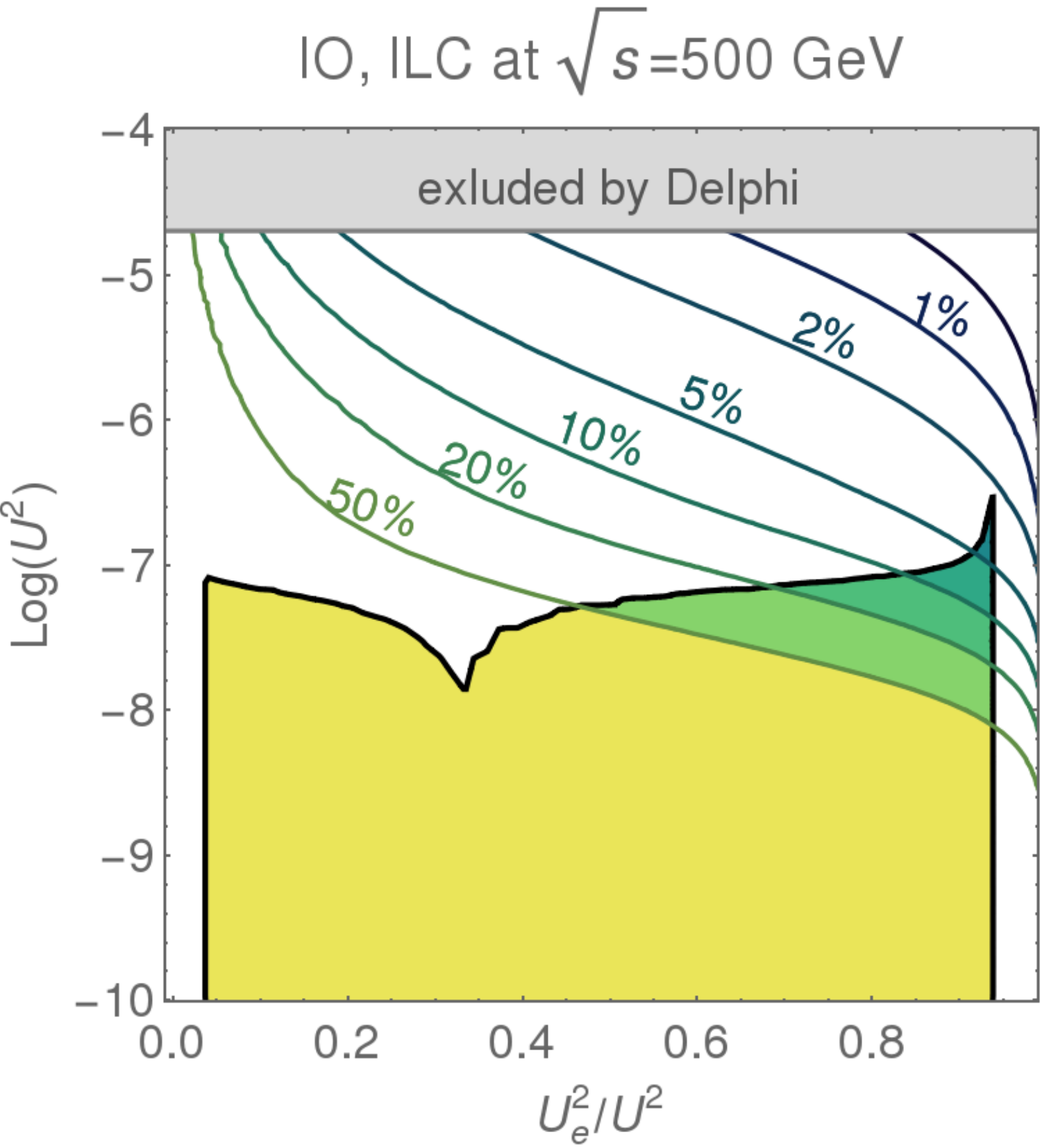}			

			&
			\includegraphics[width=0.115\textwidth]{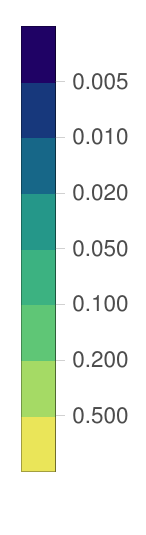}
		\end{tabular}
\caption{\label{fig:Precision-ILC}
The lines indicate the precision that can be achieved for measuring $U_e^2/U^2$ at the ILC with $\sqrt{s}=90\,\text{GeV}$ (left) and $\sqrt{s}=500\,\text{GeV}$ (right) for the case of inverse ordering (IO) of the light neutrino masses. 
The coloured area corresponds to the parameter region where leptogenesis with $n_s=2$ was found to be feasible in ref.~\cite{Antusch:2017pkq}.
 The two heavy neutrinos are assumed to be almost degenerate in mass at $\bar{M}=10\,\text{GeV}$.
}
\end{figure}

\begin{figure}
	\centering
	\begin{tabular}{cc}
		\textbf{Normal ordering} & \textbf{Inverted ordering}\\
\includegraphics[width=0.45\textwidth]{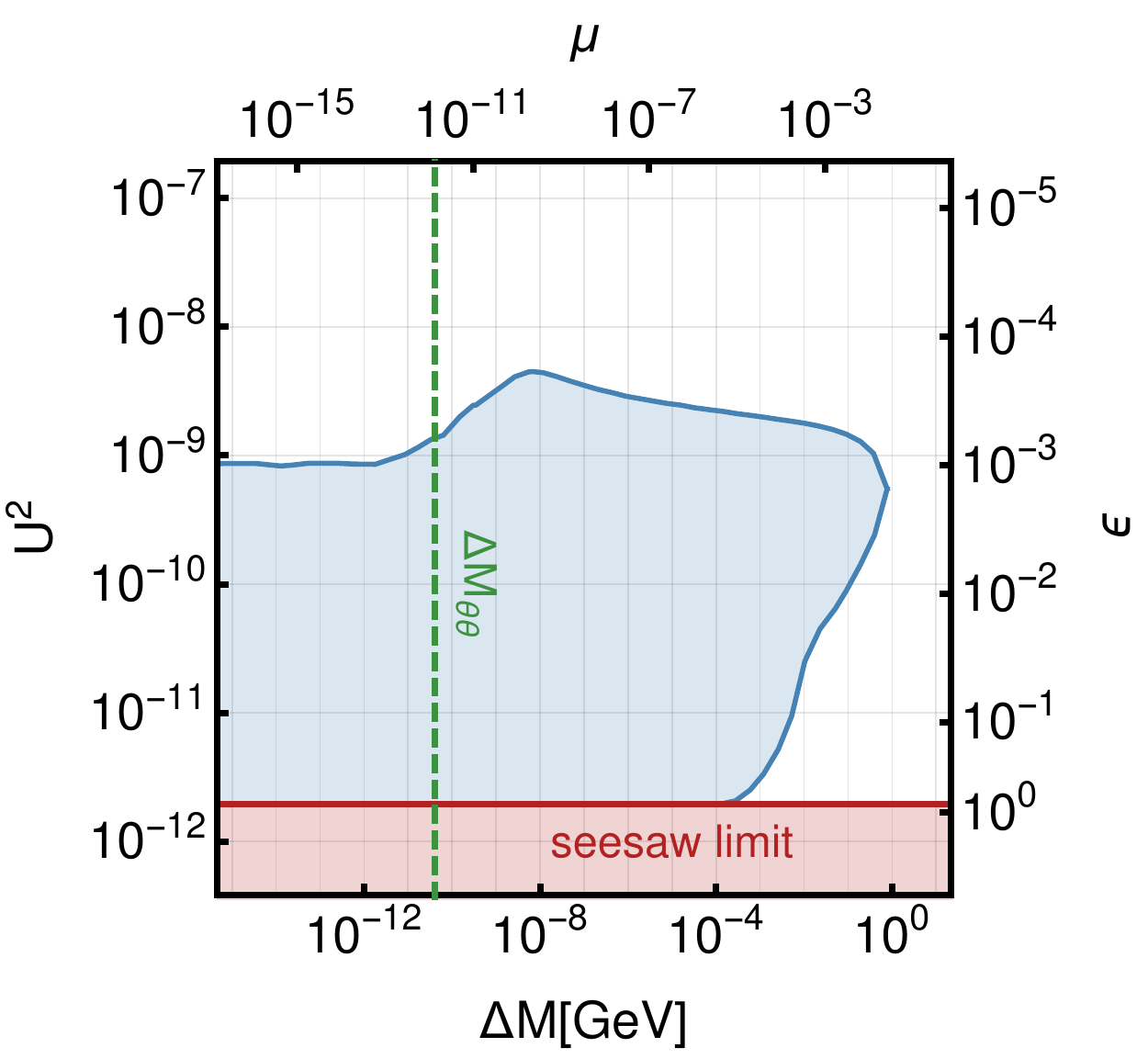}&
\includegraphics[width=0.45\textwidth]{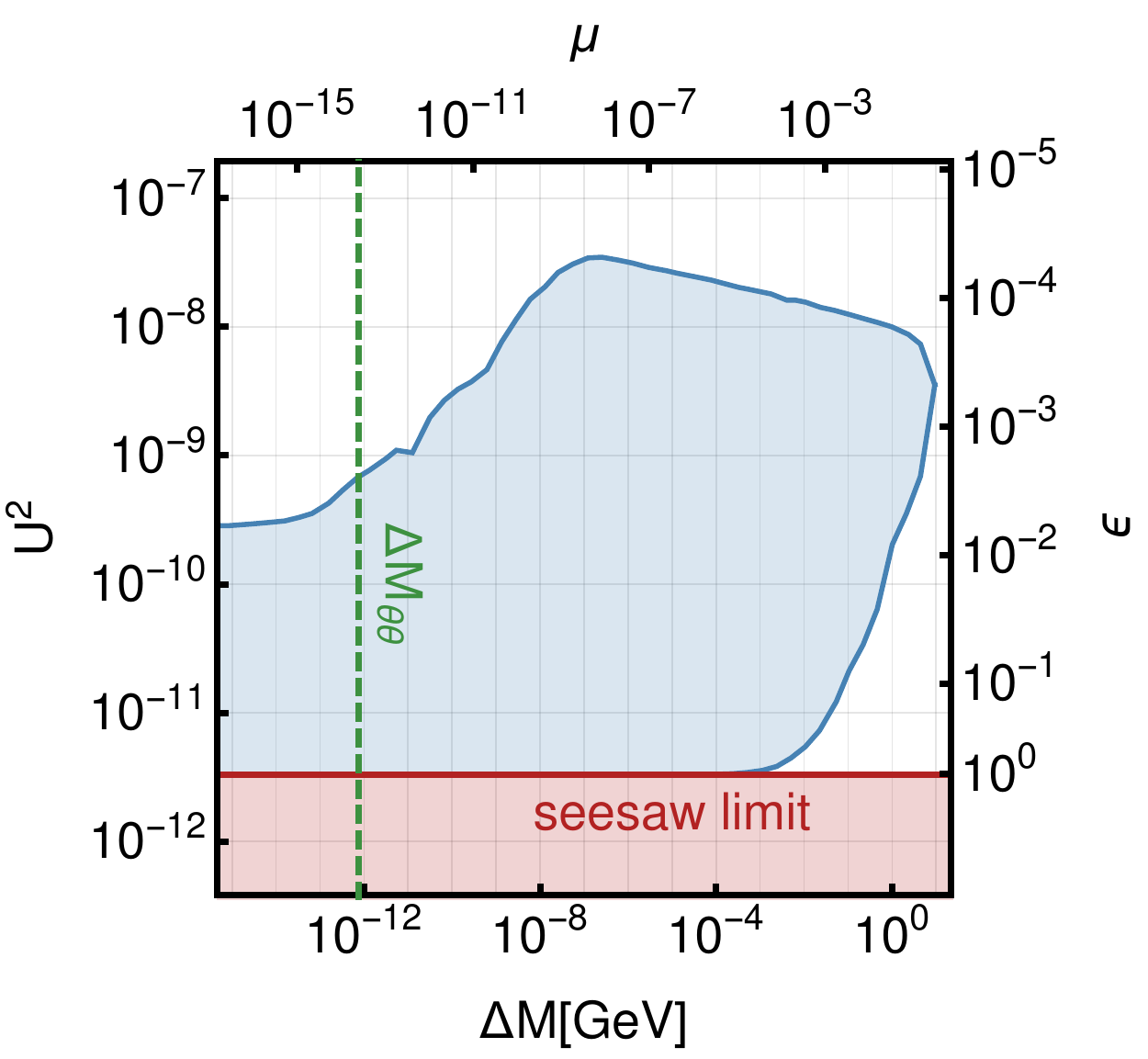}\\
\includegraphics[width=0.45\textwidth]{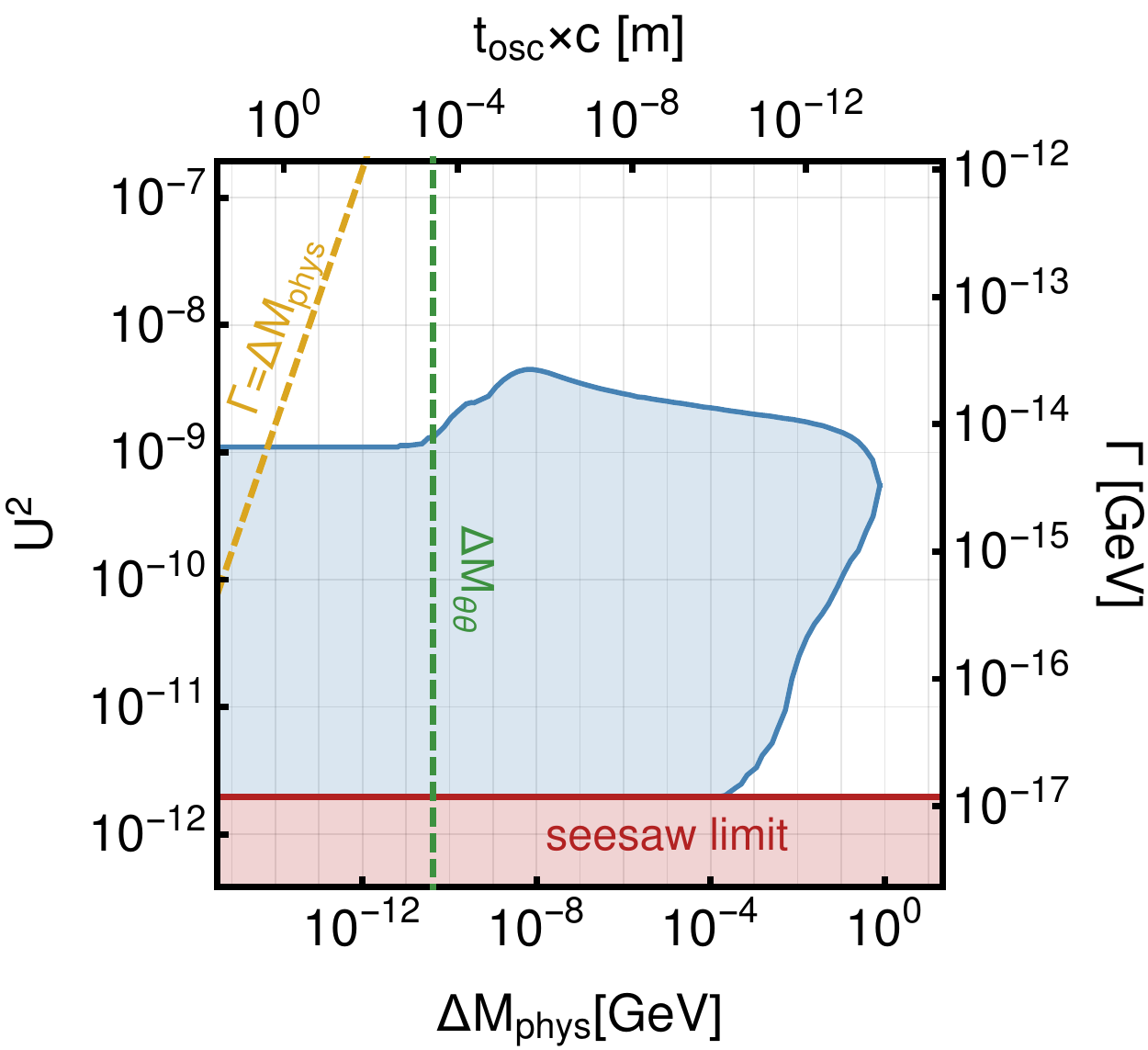}&
\includegraphics[width=0.45\textwidth]{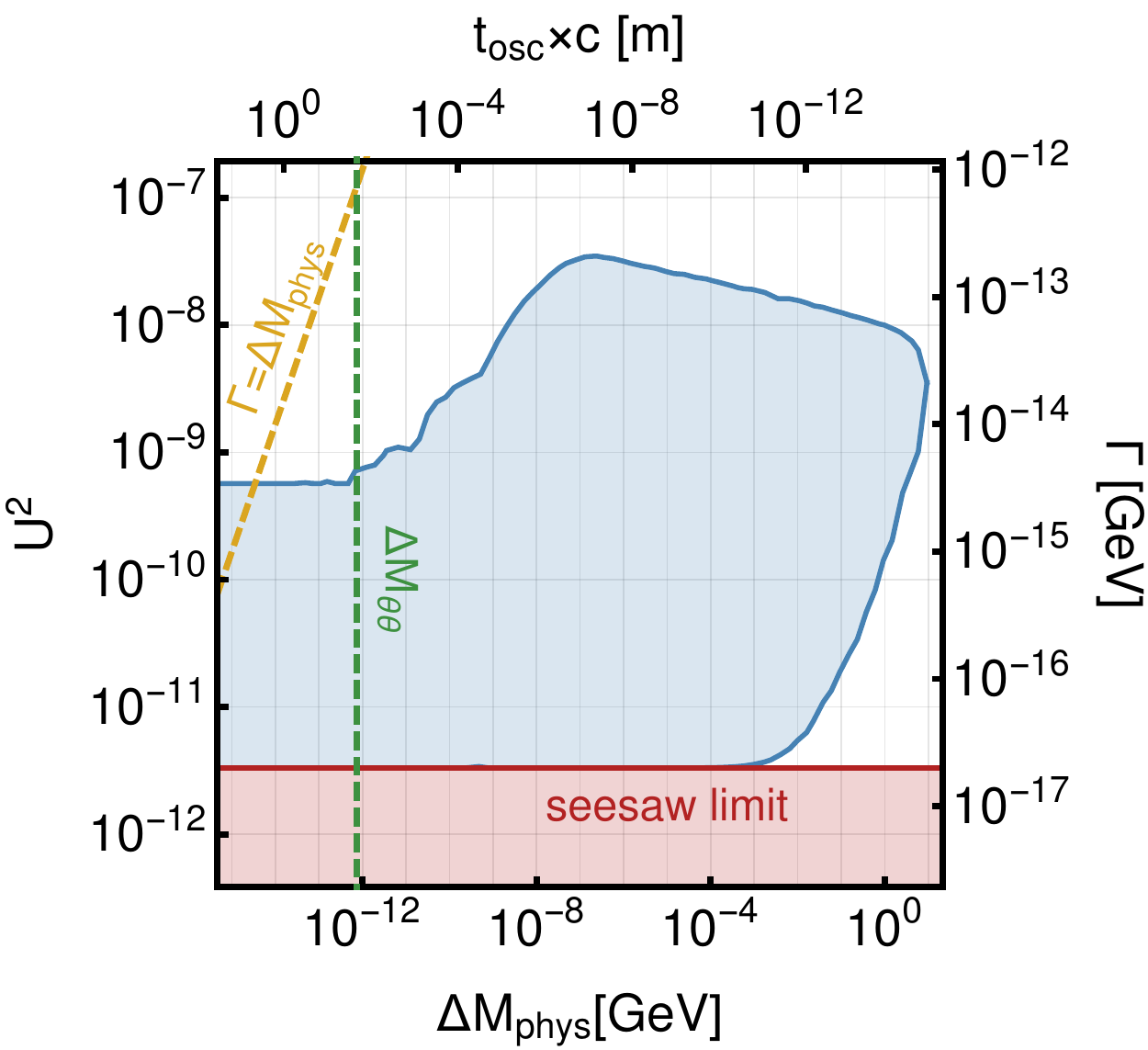}
\end{tabular}
\caption{The allowed total mixing $U^2$ in comparison to the splitting
of the eigenvalues of the Majorana mass $M$ in the Lagrangian ($\Delta M$, upper panels)
and the physical mass splittings ($\Delta M_{\rm phys}$, lower panels). $\Delta M_{\rm phys}$ is given by the eigenvalues of $M_N$ and involves corrections from the Higgs mechanism \cite{Shaposhnikov:2008pf}. It determines the oscillation time $t_{\rm osc}$ of the heavy neutrinos in the laboratory, which can be compared to their lifetime $1/\Gamma$, where $\Gamma$ is the $N_i$ decay width, cf. the yellow line.
We used an average mass $\bar{M}=30\, \GeV$.
The red line represents the "seesaw limit", below which the parameter region is excluded by neutrino oscillation data for $n_s=2$.
The vertical, dashed, green lines correspond to the contribution to $\Delta M_{\rm phys}$
 solely from the coupling to the Higgs field.
Note that leptogenesis is possible even for $\Delta M = 0$ due to this contribution during the electroweak crossover.
Figure taken from ref.~\cite{Antusch:2017pkq}.
}
\label{fig:mass_splitting}
\end{figure}

\paragraph{Testing Leptogenesis.} If any heavy neutral leptons are discovered at ILC, independent measurements of the $U_{ai}^2$ would in principle allow to determine all parameters in the Lagrangian (\ref{eq:Lagrangian}) with $n_s=2$ \cite{Drewes:2016jae}, making the minimal low scale seesaw a fully testable model of neutrino masses and baryogenesis. This may, however, be practically difficult because leptogenesis with $n_s=2$ requires a mass degeneracy $\upmu \equiv |M_2-M_1|/(M_2+M_1) <0.1$ \cite{Antusch:2017pkq}, with $\upmu\ll 0.1$ in most of the parameter space, cf. figure~\ref{fig:mass_splitting}. It may therefore not be possible to resolve the signatures of $N_1$ and $N_2$, so that the experiment is only sensitive to the combined mixings $U_a^2 = U_{a1}^2 + U_{a2}^2$. However, since $U_{a1}^2 \simeq U_{a2}^2$ in the B-L symmetric limit, this measurement already provides a strong test of the hypothesis that these particles are the origin of neutrino masses \cite{Hernandez:2016kel,Drewes:2016jae,Caputo:2016ojx} and allows to constrain the CP-violating phases in $U_\nu$ \cite{Hernandez:2016kel,Drewes:2016jae,Caputo:2016ojx}.
Such a measurement would also provide a test of leptogenesis, as not all combinations of the $U_a^2$ that are in agreement with neutrino oscillation data can lead to successful leptogenesis \cite{Hernandez:2016kel,Drewes:2016jae}, cf. figure \ref{fig:triangleplt}.
As an optimistic example, we show the precision at which $U_e^2$ can be determined with the ILC for IO in figure \ref{fig:Precision-ILC}.
However, an identification of the flavour mixing pattern alone would not be sufficient to prove that the $N_i$ are indeed responsible for baryogenesis because the BAU strongly depends on the heavy neutrino mass spectrum, and less strongly on an additional phase in $\theta$ that does not appear in $U_\nu$.
A direct kinematic measurement of $\upmu$ is only possible in a small fraction of the leptogenesis parameter region \cite{Antusch:2017pkq}, indirect measurements may be possible from a comparison of the rates for lepton number violating and conserving processes  \cite{Dib:2016wge,Anamiati:2016uxp} or $N_i$ oscillations in the detector \cite{Antusch:2017ebe}.

\paragraph{Conclusions.}
We have studied the potential of the ILC to discover heavy neutrinos that can simultaneously explain the light neutrino flavour oscillations and the origin of the baryon asymmetry of the universe.
We have focussed on the minimal model with two heavy neutrinos, which effectively also describes leptogenesis in the $\nu$MSM.
For heavy neutrino masses below the Z mass $m_Z$, the best sensitivity can be achieved with searches for displaced vertices.

We find that the ILC has the potential to observe a few hundred displaced vertex events from the decays of heavy neutrinos.
For centre-of-mass collision energies $\sqrt{s}$ at the Z pole ($\sqrt{s}=m_Z$), this number is roughly independent of the heavy neutrino flavour mixing pattern because the heavy neutrinos can be produced in the decays of on-shell Z bosons.
At $\sqrt{s} = 500$ GeV, where the heavy neutrinos are mainly produced via charged current interactions, the production relies on their mixing with the first generation, which is predicted to be suppressed for a "normal ordering" of light neutrino masses.
The performance of a Compact Linear Collider (CLIC) at $\sqrt{s} = 500$ GeV for similar values of the luminosity would be comparable to that of the ILC, while it is expected to be better (compared to the ILC's high energy run) at the higher planned collision energies because the heavy neutrino production peaks at energies above 1 TeV.

For the largest mixings $U_i^2$ consistent with leptogenesis this allows to extract information about the heavy neutrino flavour mixing pattern at a precision of a few percent. If any heavy neutral leptons are discovered at the ILC, this measurement provides a test for the hypothesis that these particles are responsible for the origin of neutrino masses.
Together with a determination of the heavy neutrino mass spectrum it can also be a first step towards probing leptogenesis as the mechanism of baryogenesis.

\paragraph{Acknowledgements.}
This research was supported by the DFG cluster of excellence 'Origin and Structure of the Universe' (www.universe-cluster.de),
by the Collaborative Research Center SFB1258 of the Deutsche Forschungsgemeinschaft,
by the Swiss National Science Foundation, by the ``Fund for promoting young academic talent'' from the University of Basel under the internal reference number DPA2354, and it has received funding from the European Unions Horizon 2020 research and innovation programme under the Marie Sklodowska-Curie grant agreement No 674896 (Elusives).

\printbibliography

\end{document}